# THE ROLE OF HOSTING CAPACITY STUDY IN POWER SYSTEM ADVANCEMENTS: A REVIEW

Utkarsh Singh

Address: Depsys SA, Route du Verney 20B, 1070 Puidoux, Switzerland

Email: utkarsh.singh@ieee.org

**Abstract:**

The fast depletion of conventional energy sources due to increased energy demands and environmental concern has motivated power utilities to integrate more renewable energy sources (RESs) into their power systems. Due to the intermittent nature and low or non-existent inertial response of these sources, a high penetration of RESs can lead to various issues in the operation of power systems such as oscillations in power system's voltage and frequency, increased harmonic distortion, failure of protective equipment, overloading of transformers and feeders, and increased line losses. Such problems arise when the hosting capacity (HC), defined as the maximum RES capacity that can be installed without having any technical and operational problems, of the network exceeds its limit due to the increased integration of RESs to the existing network. This paper reviews the progress made in HC assessment and enhancement of electrical networks research and development since its inception. Attempts are also made to highlight the current and future issues involved in HC technology for the development of an affordable, in-exhaustive, clean and reliable power supply for longer term benefits.

**List of Notations and Abbreviations:**

| | |
|---|---|
| BESS | Battery Energy Storage System |
| CSP | Concentrated Solar Power |
| C-CG | Column-and-constraint Generation |
| DG | Distributed Generator |
| DER | Distributed Energy Resources |
| DSO | Distribution System Operator |
| DVR | Dynamic Voltage Restorer |
| DSTATCOM | Distribution Static Compensator (fix the rest) |
| ES | energy storage |
| ED | economic dispatch |
| ELM | extreme learning machine |
| HC | hosting capacity |
| HCC | hosting Capacity Coefficient |
| LHC | locational hosting capacity |
| OLTC | on load tap changer |
| PV | photovoltaic |
| PSI | power supply industry |
| PVIS | photovoltaic inverter system |
| PPF | probabilistic power flow |
| PI | prediction interval |
| PSO | particle swarm optimization |
| RES | renewable energy source |
| SLHC | stream-lined hosting capacity |
| SCUC | security-constrained unit commitment |
| TSO | transmission system operator |
| UC | unit commitment |
| VaR | value-at-risk measures |
| WG | wind generator |
| WPF | wind power forecast |

## 1. INTRODUCTION

The ever-increasing concerns over global climate change, caused by the excessive use of fossil fuels to meet the global energy demands, have encouraged intensive research for green power plants with advanced technology. In this context, since past few years, the use of renewable energy sources, such as wind, tidal, micro-hydro, biomass, geothermal in general, and solar for generation of electrical energy has increased tremendously [1]-[4]. This has become possible, only because of the power system reforms. The reasons behind power sector reforms /deregulation worldwide have either been regulatory failures, political reforms, high tariffs, inefficient management, poor efficiency, or global economic crisis. A significant feature of such restructuring is to allow for competition among generators and to create market conditions in the industry, which are considered necessary to reduce the cost of energy production and distribution, eliminate certain inefficiencies, shed manpower and increase customer choice [5]-[10]. Authors in [5] have presented a detailed review on the progress of the movement to privatize and liberalize the power sector in developing countries. Authors have clearly spelled out that a full-scale power reform program generally consists of four main elements: (i) formation and approval of a power policy by the government that provides broad guidelines for the reform program (ii) development of a transparent regulatory framework for the energy market, (iii) unbundling of the electric power supply chain to enable the introduction of competition to improve sector performance in terms of efficiency, customer response, innovation, and viability, and (iv) focus on government's role on policy formation and execution, while divesting the power of state ownership at least in most of the generation and distribution. In [6], a detailed discussion on achievement of power sector deregulation in Latin America, problems encountered in the development of all three sectors of power industry i.e., generation, transmission and distribution are presented. New challenges in deregulation of electrical sectors are also presented. Authors have stressed that more emphasis should be given on the regulating conduct rather than industry structure, free access to international networks, transparent bidding process to award contracts, and the development of price signals as the base for market development and for the linkage between different segments. Enough flexibility should be provided to the regulators to make changes as and when required in agreement with the energy sector and agents of the market. Authors in [7] have thrown light on power sector reforms in the developing world. In this, authors have a detailed discussion on nine important points; (i) meaning of power sector reform and its necessity, (ii) spread of power sector reform in developing world, (iii) effect of political economy on uptake of power sector reform, (iv) works undertaken to restructure utilities and improve governance, (v) contribution made by the private power sector after reform, (vi) whether countries have established the meaningful regulatory frameworks, (vii) progress made by the wholesale power markets, (viii) improvement in efficiency and cost recovery, and (ix) outcome of power sector reform. In the last, the authors have concluded that the link between power sector reforms and final sector outcomes is much weaker, despite some evidence that private sector participation has made a positive contribution. In [8], authors have made a comparative study of the distributed and centralized technique for controlling the distribution network voltages in terms of the capacity of RES that could be integrated within the existing networks as well as contrasting them with the reactive power control approach, using optimal power flow analysis. It is concluded that both distributed and centralized voltage control methods offer significant gains in absorption capacity, particularly in rural networks. It is also inferred that the consequent losses increase substantially, and hence the financial implications of increased losses must be carefully assessed. Authors in [9] have described a cloud shadow model to recreate the variable output power of both distributed and large centralized PVs at various locations on a feeder. For this purpose, a time series load flow analysis is used using an actual EPRI test feeder. The feeder was monitored at all buses, and PV induced voltage quality was measured, including its impacts on voltage control devices. Authors [10] have presented the motivation behind RES/DG integration into the electrical network and the key issues concerning this. A brief discussion on the main challenges that must be overcome in the integration of DG into the supply systems is also provided. Authors have also emphasized on new grid code and standards to be suitable for contribution of large scale RESs.

However, there always exists a conflict of interest amongst the RES investors and distribution system operators (DSOs), as the RES owners are interested in integration of more amount of RES into the existing electrical networks, while DSOs are more concerned about the technical problems caused by excessive penetration level of RESs [11]. High penetration of RESs, predominantly wind power can lead to complexity in the operation of power systems due to the intermittent nature of these sources and frequency stability problems due to the decoupling of the RESs from AC grid using power converters [12], [13]. RESs typically have low (in case of variable speed wind turbines) or non-existent (in case of solar power plants) inertial responses. Attempts have, however been constantly made to mitigate the arising issues in RESs integration and to address various problems.

This paper deals with a state-of-the-art discussion on holding capacity of the power network, highlighting the analytical and technical considerations as well as various issues addressed in the literature towards the practical realization of this technology for better utilization of RESs, at reduced cost and high efficiency. One hundred and twenty-two publications [1]-[122] are reviewed and classified in 5 parts.

## 2. Hosting Capacity Concept

Hosting capacity (HC) is defined as the amount of RESs/DGs that can be integrated into a given power system network, while keeping its performance within an acceptable range and without having any change in the existing power system infrastructure [14]-[16]. The concept of HC was initially conceived by the computer engineers to define the capacity of a web server to host many access requests. Andre Even, for the first time, introduced this terminology for electric power applications to identify the effects of high RES into the power distribution network. The concept was later refined and clearly defined by Math Bollen and Fainan Hassan as the maximum capacity of RES that can be integrated into the system within the acceptable performance indices [11], [15], [17], [18]. Figure 1 shows an example of a HC limit [15] where a performance index is considered. With the increase in RES, an acceptable deterioration is defined. When the amount of new RES generation increases, the performance index will pass a limit, after which the deterioration is unacceptable.

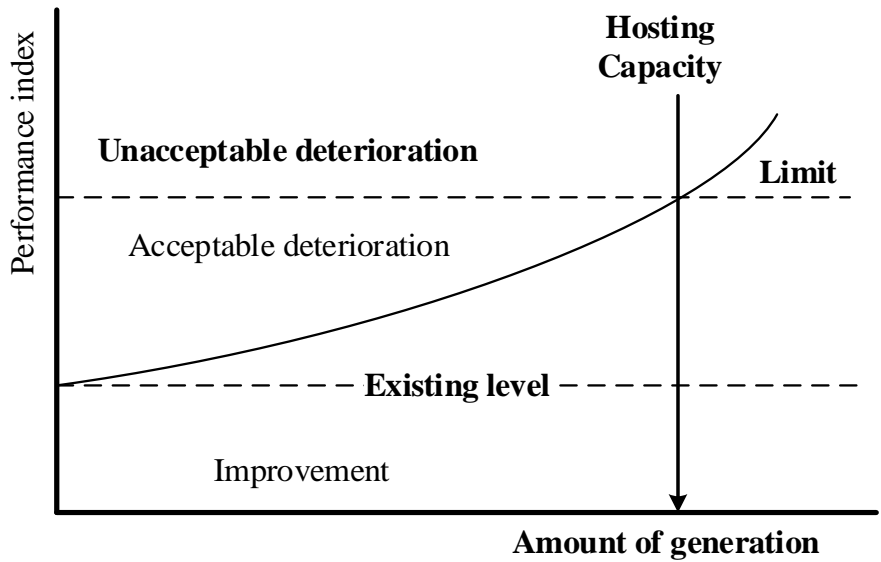

Figure 1. Hosting capacity (HC) limit and increase in RES integration [15]

Authors in [14] have presented four methods such as deterministic, stochastic, optimization-based and stream-lined to calculate HC of the distribution networks. It is shown that each method has its own advantages and disadvantages. It is also concluded that, if the generation profile is known or forecasted, deterministic method is found to be most suitable for sizing a single RES at specific location. Whereas stochastic technique can be used for sensitivity study and forecasting. In [15], HC technique is clearly explained. This approach uses the existing power system as a starting point and considers the way, in which distributed generation changes the performance of the system, without having to reinforce the operational equipment (additional conductors, exchange of transformer etc.). For this, a set of performance parameters is needed. Authors in [16] have presented the three different methods named deterministic, stochastic and time series for quantifying the solar PV HC of low voltage distribution grids. Authors have also elaborated the merits and demerits of all the three methods. It is inferred that the deterministic method is fast, and accuracy of this method depend on the model and method used to calculate the voltage rise, whereas stochastic and time series techniques require large number of simulations and computational time, which is a serious issue. In [17], three different decentralised voltage control methods, using reactive power by PV inverters, are compared for their capabilities to limit the voltage rise within a balanced low voltage system. Authors have shown that the static reactive power supply methods, as per German guidelines for generators connected to the low voltage distribution network, can increase the absorption capacity of low voltage network without having any change in the existing power network infrastructure.

Authors in [18] have applied HC approach to a real network in order to calculate the absorption capacity for integration of new RES. In this case study, two limits overvoltage and overcurrent setting the HC were evaluated. The important findings of these works [11], [14]-[18] are summarized as:

- For a given distribution network, there is no single value for HC, as it depends upon the pre-defined limiting factors to calculate it.

- Inclusion of so many limiting parameters makes the HC analysis extremely complicated.

- It is difficult to point at a single method for quantifying the HC as the most suitable one. It requires an exhaustive application of the methods to many low-voltage distribution network and a qualitative comparison of the results.

- Deterministic methods use traditional power flow analysis and assume that inputs are fixed and known. This technique neither considers the varying power production due to the change in wind speed in case of wind energy and irradiation in case of solar energy nor varying consumers' power consumption.

- Stochastic and time series methods are more suitable because both methods include variable input parameters i.e., uncertainties due to change in wind and solar power production and customers' power consumption in the analysis.

- In case of stochastic method, computational burden is more due to the inclusion of uncertainties in customer consumption, grid and wind and solar power generation.

- Time series method is better than stochastic method as it considers all the time-dependencies and correlations. This method can be successfully applied for time varying assessment of hosting capacity and the response of protection /control devices.

- If the location, size and number of RESs is known, optimization method will prove to be the best choice, since it can improve overall performance of the distribution network by reducing the losses and/ or cost.

- The selection of HC calculation method will always depend on the objective of the study.

- Any HC study needs three major inputs: performance index, a corresponding limit, and a method to calculate the performance index as a function of new power production or consumption.

## 3. Impact of Wind Energy Integration

RES technologies are not yet economically competitive with conventional thermal generation, as high-power penetration impacts the power supply industry (PSI) technically and economically both. From the technical point of view, PSI faces a

7variety of problems and challenges such as frequency and voltage regulation, power quality issues, available transmission and distribution capacities to accommodate RES plants, monitoring and control, operational practices, ancillary services, connection interfaces, etc. Large wind generator (WG) integration will also impact energy balances and generation mix, electricity markets, and emissions, etc. [18]-[22]. Most of the technical challenges are related to volatile nature of wind. Wind power generation fluctuates based on wind speed, which depends on the geographical location. Since the amount of wind may vary from time to time, the variable wind power causes other conventional generator to operate in sub-optimal manner [23]. Power systems must incorporate, for the first time, a source of high uncertainty, high volatility and low predictability [24]. This uncertainty includes input data, power curve, weather conditions and prediction algorithms. Forecasting of wind generation is very challenging and extensive research has been done on this topic [25]-[28]. Authors in [19] have detailed discussion on the impact of high wind power integration on PSI, technical challenges and solutions required. Apart from the technical issues, large wind power integration will have severe impact on PSI economics and affects the other market participants. Authors have stressed that a detailed analysis should be carried out on the important issues like inter-TSOs energy trading, impact on generation mix, energy cost, energy balance, reliability and security of supply. In conclusion, authors have recommended for the design and development of new grid codes and market regulations to ensure the security of supply in terms of system security and generation capacity adequacy. In [20], authors have reported that the integration of RES will change the generation mix against conventional power plants and will also reduce their market share. This impact may be significant for the new market entrants. Due to this, targets for market opening and large-scale RES integration may conflict each other. Therefore, policies and regulations should be carefully designed to benefit both conventional and RES sectors to mutually achieve the said goals. Works reported in [21] is about fault ride through, grid voltage control, system monitoring and protection as well as retrofitting of old units. In the same paper, authors have discussed the new requirements, defined in accordance with the new developments in wind turbine technologies, which should be utilized in future to meet the grid requirement. Monitoring and system protection is defined under the aspect of sustainability of the measures introduced. In [22], authors have developed a new control scheme for variable speed wind turbine that enables power supply of islanded parts of a distribution network. Models and strategies for fast control of frequency and voltage during islanding are also derived. Based on the simulation results, authors have concluded that wind turbine can operate as conventional generator units supplying power to islanded parts of the distribution grid, while maintain the voltage and frequency close to their normal values. It is further inferred that energy storage (ES) is not required for power balance control, as the frequency loop of the control scheme is very fast, ensuring stable operation in islanded mode. Authors in [23] have presented the wind power variability and its impacts on power systems, covering classification of variability, aggregation effect, and wind power forecasting error. The effects of wind energy on different operational time fames of conventional power plants, and the cost of balancing requirements to accommodate high wind power integration levels are also discussed in this paper. Authors have concluded that though the wind power balancing cost is highly system dependent, but in general, it increases with the increase in penetration level. The flexibility of existing dispatchable generation units and the available transmission capacity to neighbouring areas play an important role in reducing the balancing costs. Authors in [24] have reported a detailed discussion on integrating wind energy into the electric power system. This includes topics like commercial wind generation technology, development of wind technology, reactive power supply and voltage control, active power regulation, frequency control, wind forecasting, forecast accuracy, effect of spatial spread, and power balancing issues. It is concluded that the stalemate in transmission development is coming to an end, with the new transmission planning paradigm being implemented. Study embodied in [25] presents the detailed investigation on wind power forecast (WPF) uncertainty in unit commitment (UC) and economic dispatch (ED) and analyses the impact of different reserve requirements and UC policies on system operations. It is demonstrated that despite the inherent uncertainty and risks, it is possible to adopt a quantified method to deal accurately with wind power generation. Authors conclude that stochastic decision methods can be successfully adopted to reduce the costs and risks, in the presence of large and unavoidable wind power prediction errors, particularly in power system network dominated by thermal power plants. Authors have further suggested for the use of adaptive reserve requirements, which are the function of wind power forecast. Paper [26] presents the impact of wind forecast error statistics upon unit commitment for a large wind integration test system. Authors further report that variance has the most impact and suggest that if skewness is included in the evaluation of error information, kurtosis should also be included to reduce the system cost. In the same work, it is inferred that the interactions of variance, skewness, and kurtosis changes the utilization and commitment of units, and the representation of variance, skewness and kurtosis can affect the dependency of commitment upon flexible units and the way it is used. Researchers in [27] presents a state-of-the-art review on probabilistic forecasting of wind power generation and a brief discussion on three different representations of wind power uncertainty. Three different forecasting methods, in terms of uncertainty representation, i.e., probabilistic forecasts (parametric and non-parametric), risk index forecasts and space-time scenario forecasts have also been discussed. In the last, authors have concluded that uncertainty forecasting techniques can reduce the economic and technical risk caused by wind power uncertainty and have recommended that there is need of more research on wind power uncertainty forecasting. A new hybrid intelligent algorithm approach to produce prediction intervals (PIs) of wind power generation, based on the extreme learning machine (ELM) and particle swarm optimization (PSO) is presented in [28]. Optimal PIs have been obtained without prior knowledge, statistical inference or distribution assumption of forecasting errors that are required in most of the traditional probabilistic techniques. It is concluded that the proposed hybrid scheme provides a general framework of probabilistic wind power forecasting with high flexibility for reserve determination by transmission system operator (TSO) to meet the load, and to economically operate the systems. A two-stage stochastic unit commitment model integrated with value-at-risk measures (VaR), including non-generation resources like demand response and energy storage systems, to balance between cost and system reliability due to the fluctuation in variable power generation is presented in [29]. To solve the VaR-based model and to reduce the computational time, modified Benders' decomposition algorithm is used. Sensitivity analysis is carried out for evaluation of reliability parameters to reduce the generation costs. Numerical experimentation is also carried out to find optimal unit commitment solutions and to compare the effect of risk of non-generation resources on power generation. Li et. al. [30] have proposed a stochastic dynamic model to formulate the spatial and temporal correlation between the atmospheric and near-surface wind fields of geographically distributed wind farms. It is inferred that the model can provide competitive interval forecasts with conventional statistical based models. In [31], a robust optimization-based analysis is carried out by combining the concentrated solar power (CSP) plants with wind farms to reduce the overall uncertainty in the joint power output. A new approach, based on nested column-and-constraint generation (C-CG) method, is developed to solve the multilevel optimization with mixed-integer recourse problem. The proposed technique is found to be suitable in identifying robust yet narrow intervals in a reasonable amount of time. It is concluded that the combination of different types of renewable resources is very effective in reducing the generation uncertainty. Hu et. al [32] have thrown light on how variant robust security-constrained unit commitment (SCUC) models in terms of different worst-case definitions could impact operational security and economics of power systems under uncertainties. In addition, what is the proper robust SCUC model that can fulfil the specific market operational requirements of independent system operators /regional transmission owners for effectively operating the system. Four different robust SCUC models are studied by the authors. In [33], various techniques for inertia and frequency control developed for variable speed wind turbine and solar PV generators are systematically reviewed. Authors [34] have presented a more tractable and adaptive distributionally robust unit commitment (DRUC-dW) formulation using distance-based data aggregation, and an efficient cutting plane algorithm, which solves the two-stage problem quite efficiently by leveraging the extremal distributions constructed. It was found that UC solution, yielded from DRUC-dW model, achieves a reasonable balance between the robustness and cost efficiency. In [35], a non-anticipative robust unit commitment models (NRUCs), where determining the dispatch policy is delayed until the uncertainty decreases is developed. The proposed NRUC features three decision-making problems sequentially solved under different degrees of uncertainty. At first, the decision-making problem is formulated as an intractable three-stage robust optimization problem. Then, a suboptimal approach is developed, where a constraint is imposed on the dispatch policy so that the transmission capacity constraint is met irrespective of the dispatch level. The significant findings of the studies may be summarized as follows [18]-[35]:

- Wind turbine generators are not equipped with exciter and voltage control and have no ability to provide high short-circuit currents to reduce the voltage dip. Hence, if voltage drops below a certain level, WGs trip.

- The impact of wind power integration on any power network depends on the penetration level and system flexibility. Increase in the penetration level increases the impacts perceived by the power network, whereas system having more flexibility can accommodate more power without perceiving unwanted effects caused by wind generation units.

- Impacts of wind power integration are classified as short-term and long-term impact. Short-term impacts deal with operational time scale, such as system balance issues, which are represented by the requirements and cost caused by the fluctuating wind power. Long-term impacts are related to planning for peak load periods.

- Fluctuating wind power causes the other conventional power units to operate in sub-optimal manner with reduced efficiency. This problem can be reduced by accurate wind generation forecasting and prediction.

- Depending on the penetration level, wind power can increase or decrease system (transmission and distribution) losses.

- Fluctuation in power output leads to voltage variation in case of fixed-speed wind turbine. However, the variable speed turbines, such as doubly fed induction generators, can provide reactive power to network.

- As for as the impact of wind power integration on distribution network under normal operation is concerned, there may appear slow voltage variations due to change in power flows and short duration voltage dips during switching (on/off), fast voltage variations due to changes in wind speed that may cause 'flickering' effects, and voltage distortion due to harmonics.

- During network faults, integration of WG can lead to: increased stress of circuit breakers since the short circuits are additionally fed by the wind generators, malfunctioning of protection equipment, since it is designed to operate within strictly 'radial structure' of the MV network, and islanding of parts of the grid, fed solely by wind generators, which may cause failure of the end-use equipment and also accidents to the utility 's personnel. It is more possible in the presence of high capacitance equipment such as cables.

- High penetration of wind power needs expansion of more complex transmission networks resulting in more transmission losses.

- The long distance between the wind forms and load requires long transmission lines to be laid down and can lead to transmission congestion.

- With the increase in wind power integration, the amount of reserve needs to be increased over long period of times to maintain the system reliability.

- High integration of wind power leads to frequency deviation from the normal range, because maintaining the balance between supply and load demand in case of high volatility of supply is challenging and more difficult.

- The increase in penetration level of wind power results in an equivalent decrease in conventional generation units, and thus system rotational inertia becomes very low. This can lead to serious effects on the frequency deviation.

### 4. Impact of Solar Energy Integration

Solar power is the conversion of sunlight into electricity, either directly using photovoltaic (PV) or indirectly through concentrated solar power (CSP). Solar power in its various forms such as solar photovoltaic, solar heat, solar thermal electricity and solar fuels offer a clean, climate-friendly, very abundant, and inexhaustive energy source to mankind [36]. Solar power generation is one of the most advanced technologies for renewable energy production. Solar energy has delivered more new capacities than nuclear and fossil fuels [37]. Worldwide growth of solar power is extremely dynamic and varies strongly by country. By the end of 2019, a cumulative amount of 629 GW of solar power was installed throughout the world [38]. By early 2020, the leading country for solar power was China with 208 GW [39, 40], accounting for one-third of global installed solar capacity. As of 2020, there are at least 37 countries around the world with a cumulative PV capacity of more than one GW. The countries in the list of top installers of 2016 through 2019 were China, United States, and India [41, 42]. The top 10 countries by added solar PV capacity in 2019 [43] and Top 10 countries by cumulative solar PV capacity in 2019 is shown [44]. The solar PV capacity by country and territory (MW) and share of total electricity consumption is given [45]-[48].

Solar-grid integration is another important technology as it optimizes the energy balance leading to improvement in economics of PV system, reduction in operating cost, and provides added value to utility and consumers. Solar-grid integration technology consists of advanced inverter technology, grid-plant protection technology, anti-islanding technology, smart grid technology and solar-grid forecasting technology [49], [50]. Authors in [49] have presented a technical report on 'Treatment of solar generation in electric utility resource planning', jointly prepared by the National Renewable Energy Laboratory (NREL) and Solar Electric Power Association (SEPA). A detailed study was conducted on inclusion of solar in long-term resource planning processes through interviews and a questionnaire. Table 1 includes the benefits and challenges of solar, based on utility interviews. With these benefits and challenges in mind, utilities can more accurately incorporate solar generation into their long-term planning processes. Some of the leading, utility-identified best-practices and analysis that needs attention are: (i) analyse and assign appropriate capacity values to solar resources, (ii) analyse solar individually, to get more accurate aggregate results, (iii) improve modelling assumptions and methods, (iv) pursue sub-hourly sensitivities, (v) evaluate whether to treat distributed generation as a resource, and (vi) utility-identified analysis needs. In [50], a detailed study on the effect and challenges of integration, the current solar-grid integration technology, its benefit, solar system characteristic for integration, and issues and compatibility of both the systems have been carried out by Nwaigwe et al. Authors further conclude that solar power integration can reduce the transmission and distribution losses, increase grid resilience, lower generation cost, and reduce requirements to invest in new utility generation capacity.

Table 1 Benefits and challenges of solar power integration [49]

| Benefits of solar power integration | Challenges of solar power integration |
|---|---|
| • Meet renewable standard requirements | • Variable and uncertain output |
| • Fuel diversification | • Ramping issues |
| • Cost stability | • Economics |
| • Geographic dispersal benefits and modularity | • Lack of current capacity need |
| • Partial correlation with peak demand | • Cross-subsidization concerns (DG) |
| • Mitigation of environmental compliance risks | • Reduced capacity benefit over time with increasing solar penetration |
| • Avoid line losses (typically DG only) | |

Photovoltaic (PV) technology has presently become a significant form of power generation on many power system networks. Impact of high PV integration has drawn attention towards the issues of grid management, operation and planning, particularly where there is variability in PV system output due to cloud cover. Variability in PV irradiance is considered as a major challenge to high levels of PV penetration into the existing power networks, and this variability in PV generation can have a negative impact on the local electricity network. Network level problems occur, where intermittent changes in PV generation on a power network are unable to be accommodated by the base load generation over the time frame of the change. The potential impacts of intermittent generation occur at different time scales, and the associated considerations include [51]: (i) rapid changes in network demand or in PV generation can lead to power quality issues such voltage flicker and harmonic distortion, (ii) the spinning reserve of system generation is required to have sufficient total capacity and ramping capability to meet the short term changes in network demand or in supporting PV generation, otherwise it may lead to power quality and system outage issues, and (iii) generation planning requires sufficient generation to be available at any time to fulfil the projected network demand and sufficient operating reserve.

The integration of distributed generation (DG), particularly solar power can significantly impact the flow of power and voltage conditions at customers and utility equipment. Depending on the nature of DG and operating characteristics of distribution system, these impacts may be either positive or negative. The positive impacts [52] are:

- Voltage support and improved power quality.
- Loss reduction.
- Improved utility system reliability.
- Transmission and distribution capacity release.
- Deferments of new or upgraded transmission and distribution infrastructure.
- Reduction of emissions.

The above listed benefits are, in practice, much difficult to achieve. The DG sources must be reliable, dispatchable, of the proper size and installed at the proper locations and must also meet various other operating criteria. Since many DGs will not be utility owned or due to intermittent nature of solar and wind power, it is difficult to satisfy all these requirements. In fact, power system operations may be adversely affected by the integration of DG, if certain minimum standards for control, installation and placement are not maintained. The major negative impacts are: voltage variation and unbalance, current and voltage harmonics, stress on transformer, grid-islanding and power quality issues [53], [54]. The severity of these impacts depends on the penetration level of PV, its location and configuration of distribution network. The negative impacts can be summarized [55] as:

- Voltage fluctuation in feeder, resulting in voltage rise or fall and unbalanced voltage
- Malfunctioning of on-load tap changer, line voltage regulators and capacitor banks.
- Possibility of overload in distribution feeders.
- Variation of reactive power flow caused by malfunctioning of capacitor bank devices.

- Malfunctioning of overcurrent and overvoltage devices.
- Islanding operation and detection in case of grid disconnection
- Reliability and security of the distribution networks.

The electric grid -an interconnected power network shown in Figure 2 maintains an instantaneous balance between the supply and demand (generation and load), while transferring electricity from generation source to customer [56]. If the power produced by the DER (solar PV) is more than power consumed by the load, voltage level of that load bus increases, and can cause damage to various equipment connected to that feeder. On the other hand, there will be drop in voltage, if power demand by the load is more than the power produced. In [57], authors have presented a detailed study on voltage fluctuation in power networks with photovoltaic energy sources, and a flickermeter model has been used for the evaluation of flicker assessment under climate change (sunny and cloudy situations). In the last, authors conclude that the flickermeter model fulfils the requirements defined in IEC 61000-4-15 standard, and it has been tested under additional tests defined in CIGRE/CIRED/UIE test protocol. Authors have also recommended the use of flickermeter for evaluating the flicker severity and voltage fluctuations produced by the photovoltaic energy sources. The voltage unbalance is a major power quality problem in low voltage distribution networks due to the random location and rating of single-phase rooftop photovoltaic cells (PV). In [58], Wang et al have presented a detailed study on voltage issues at the point of common coupling caused by the integration of photovoltaic system. For this purpose, a 7.2 kW grid connected PV system on a radial LV distribution network has been set up, and the probability density of voltage rise, and voltage unbalanced factors were derived from the test data. Short-term and long-term voltage flicker indexes were also calculated to evaluate the severity of flicker. It is stated that the intermittent PV power output can introduce a maximum short-term flicker of 2.0 and long-term flickers of 0.78. It is also shown that 1% of the voltage unbalance can produce 6-10 times current unbalance. To improve voltage unbalance, authors in [59] have proposed a system with converter topology and control algorithm, based on the application of series dynamic voltage restorer (DVR) and parallel distribution static compensator (DSTATCOM) custom power devices. A state feedback control, based on pole-shift technique, has been developed to regulate the DSTATCOM and DVR converters output voltage for voltage balancing in the network.

Integration of PV systems in the distribution network, harmonic distortion of voltage and current waveform takes place, resulting in poor power quality. PV inverters are the main source of harmonic injection. The maximum penetration level of grid connected identical photovoltaic inverter system (PVIS) that can be installed is based on the acceptable voltage distortion levels within the distribution network as determined in [60]. To mitigate harmonics in the network, Inductor-Capacitor-Inductor (LCL) filters with robust control techniques are currently used. For new interface converters, their EMI/EMC related impact should be considered since its early stage of design. It is not enough to mitigate the problem in final design, but to go to the root of the problem so that harmful emission does not occur [61]. New more complex converter topologies and modulation schemes are being studied to solve the power quality issues, such as harmonics, voltage imbalances etc. [62], [63].

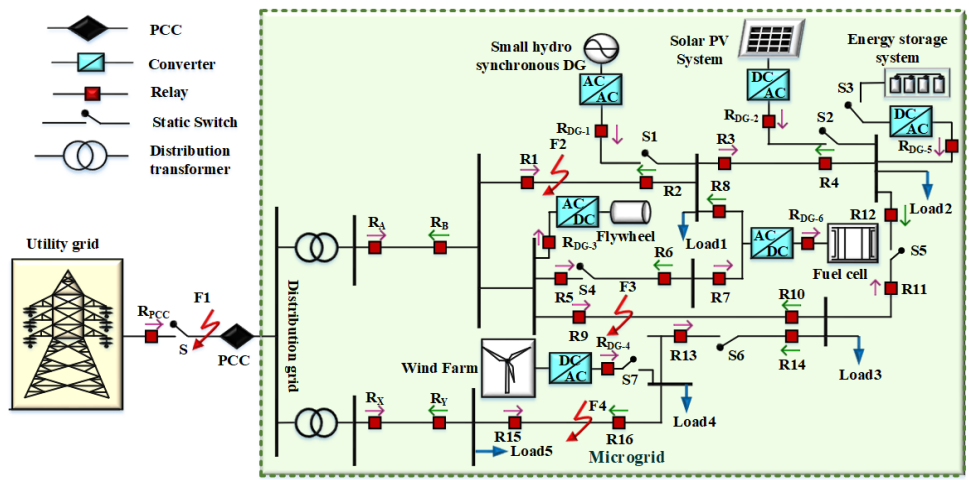

Figure 2. Structure of an AC microgrid [56], showing multiple generation units and load. To maintain a demand-supply balance and reliable operation in such a hosting capacity analysis is crucial.

The major findings of the studies may be summarized as follows [36]-[63]:

- Some renewable energy technologies provide power only when the resource is available. These resources are often contracted as 'must-take' generators, where their output is always used, when it is available. However, it is difficult to integrate a large amount of 'must-take' generation into the grid, because its availability is uncertain and constantly changing.

- Photovoltaics (PV) may be centrally located in large plants or distributed on rooftops. Distributed PV has benefits, such as low land use and no transmission needs. Distributed and central PV both are usually 'must-take' generators.

- Storing large amounts of power is difficult, while storing thermal energy is relatively easy. Since, concentrated solar power (CSP) plants collect and convert thermal energy into electricity, they can collect and store thermal energy for later conversion into electricity. CSP plants with thermal energy storage provide assurance that the generator will be available, when needed. These CSP plants are dispatchable and can meet the intermediate and baseload demands.

- Although PV deployment depends on various integration issues, most CSP plants respond more slowly to changing weather, particularly equipped with thermal energy storage system, and output from these plants is easier to forecast and integrate into the electric grid.

- PV generation without energy storage system does not provide all the characteristics required for stable grid operation.

- Careful integration of distributed generation and deployment of utility-scale generation will be needed to provide the mix of power and reliability required for a clean and healthy power supply as renewables contribute an increasingly larger share of energy needs.

- Static synchronous compensator (STATCOM) in conjunction with energy storage system is likely to play an effective role in mitigating the voltage issues without curtailing any renewable energy and can provide active power support during the grid contingency.

- Distribution system designs and operating practices are normally based on radial power flows, and this creates a special challenge to the successful introduction of distributed generation. Some important issues such as voltage regulation and losses, voltage flicker, harmonics etc. must be considered to ensure that DG will not degrade the distribution system power quality, safety or reliability.

- The fault contribution from a single small DG unit is not large, however the aggregate contributions of many small units, or a few large units, can alter the short circuit levels enough to cause fuse-breaker miscoordination.

- Distributed generation must be applied with a transformer configuration and grounding arrangement compatible with the utility system to which it is to be connected. Otherwise, voltage swells and over voltages may be imposed on the utility system that may damage the utility or customer equipment.

- Islanding can occur only if the generator(s) can self-excite and sustain the load in the islanded section. In most cases, it is not desirable for a DG to island with any part of the utility system, because this can lead to safety and power quality problems that will affect the utility system and load both.

- The implementation of DG can increase the reliability of electric service if units are configured to provide 'backup- islands' during upstream utility source outages. To be effective, this requires reliable DG units and careful coordination of utility sectionalizing and protection equipment.

- To avoid costly impact studies for all but those applications that actually need them, proposed DG can be screened based on factors such as: utility system fault levels at the point of DG interconnection, size of DG unit, its intended mode of operation and expected output fluctuations, aggregate capacity of DG, secondary configuration of the DG site (including presence of adjacent customers), feeder voltage regulation practice, transformer and grounding compatibility of the system, size of generation relative to load at the interconnection point and type of interfacing power converter.

- The main cause of new solar panel failure is poor design and defects during manufacturing, contact defects in junction boxes, glass breakage, burst frames, breakage of cell interconnections and problems with the diode associated with a higher rate of cell degradations and interconnectors, fuse boxes, charge controllers and cabling as well as issues with grounding.

- The variance created by the installation of a further dispersed PV inputs into power grids can end up being very similar to step-change 'noise' variance, which currently occurs in the network.

- Analysis of the size, number, and spatial diversity to optimise PV input into the grid should be undertaken with a view to determining the marginal benefit of additional diversity and/or the extent to which the benefits of diversity diminish, if the separation of systems gets too great.

- Under-generation and over-generation both by solar PV cause instability on the grid. Effective solution to this involves use of better forecasting tools for more accurate prediction of when solar generation might decline to minimum penetration capacity, installing solar panels across a large geographic area to minimize any impact of generation variability due to local cloud cover, shifting power supply and storing excess energy for later use and encouraging customers to use power, when it is more readily available.

- Prior to deployment of solar panels, advanced integration technologies should be considered. Optimized forecasting is essential for proper system stability. Also, due to economic viability and robustness of the system, solar technology can be treated as a major guideline for sustainable development.

- Integrating CSP plants with wind power can reduce the generation uncertainty and improve the capacity factor of the combined plant, because of the negative correlation of wind and solar power, and the availability of high-efficiency thermal energy storage systems.

### 5. Assessment of Hosting Capacity (HC)

The HC concept has been most widely used to evaluate the benefits of different voltage regulation techniques, amount of solar and wind energy that a grid can accommodate without violating the acceptable performance indices of the existing power network. Various power system phenomena and related performance indices are given in [64], and the same is reproduced in Table 2. Distributed generation (DG) such as solar and wind will impact the performance of the grid, and this sets a limit to the amount of such RES that can be integrated. New communication schemes together with energy storage system and grid control strategies allow integration of increased amounts of renewables into existing power networks, without unacceptable effects on users and grid performance. The first work on the hosting capacity of electrical grids was reported by Math Bollen and Hager in 2004 [65]. Later, many methods have been applied to examine the capacity of existing distribution grids to accept DG [66]. However, some important differences do exist between methods. Most of the statistical approaches proposed in the literature aim at defining the optimal DG location and sizing [67]. It is very essential to define suitable performance indices before calculating the amount of DG that can be integrated into the existing network, as the hosting capacity is based on this. The choice of performance indices and limit have a big influence on the amount of distributed generation that can be accepted. The hosting capacity is a tool that provides trustworthy and secure platform for fair and open discussion between the different stakeholders (network operators, owner of DG, owner of the existing large power plants and other customers) and a transparent balancing between their interests, for example, acceptable reliability and voltage quality for all customers, no unreasonable barriers against new generation and acceptable costs for the network operator [68], [69]. The integration of DG should not result in a deterioration of the supply reliability, but certain deterioration in quality of the supply should be acceptable for most customers. Reliability is quantified by several indices. The steps to be followed to calculate HC is summarized below [15]:

i. Choose a phenomenon and one or more performance indices,
ii. Determine a suitable limit or limits,
iii. Calculate the performance indices as a function of the amount of generation,
iv. Obtain the hosting capacity.

Table 2

Power system phenomena and related performance indices [64]

| S.N. | Phenomena | Performance Indices |
|---|---|---|
| 1 | Overloading from wind power | Maximum hourly value of current through transformer |
| 2 | Frequency variation | 99% interval of 3 s average of frequency |
| 3 | Overvoltage from roof top solar photovoltaic cells | Highest 10 min average of voltage |
| 4 | Undervoltage from fast charging of electric vehicles | Lowest 10 min. average of voltage |

| 5 | Protection mal-trip | Lowest recorded current causing interruption |
|---|---|---|
| 6 | Harmonics | 10 min. average of voltage and currents |

In [70], authors have presented the hosting capacity approach and explained some important developments, such as uncertainty in location and size of production units, and curtailment to connect more production than according to the initial hosting capacity. Authors have given the new HC terminology as:

  i. HC uncertainty,
 ii. HC Coefficient,
iii. Stochastic HC (SHC),
 iv. Locational HC (LHC).

Uncertainty in the estimation of HC may occur due to volatile and intermittent nature of DG output powers, unknown rating of DG units and their locations, load variability and insufficient data, when performing power system calculation. Hence, HC will never be a single value, but multiple values will be obtained depending on the uncertainty percentage. Since randomness is involved in the calculation of HC, it cannot be termed as deterministic approach, instead it should be viewed as a probabilistic approach, where level of accuracy and uncertainty are considered. Basically, HC is location-based concept; the maximum HC occurs at the maximum loading conditions and minimum generation, whereas minimum HC is obtained at minimum loading conditions and maximum generation [11], [71].

Curtailment of power production is defined as reducing the active power output from certain energy resources at times to increase the HC. The ability to curtail the production will allow for a larger installed capacity of distributed generation. The increase in installed capacity for a given percentage of permitted overloading varies with the type of energy source. The gain in hosting capacity will depend on characteristics of the network and performance index being considered [72]. This approach can be extended to cover different voltage levels and hosting capacity limits determined by very different performance indices (and therefore different power system phenomena). For an objective comparison, the Hosting Capacity Coefficient (HCC) is defined as the ratio between the curtailed energy and the installed capacity above the initial HC as given in Eq. (1). In the same work, it is inferred that a low HCC corresponds to an unfavourable location for curtailment. A high value of HCC corresponds to a high ability to accommodate DG using curtailment. However, it is possible that a high coefficient corresponds to a low HC limit.

$$\text{HCC} = \frac{\text{Curtailed Energy}}{\text{Installed Capacity} - \text{Initial HC}} \qquad (1)$$

Integration of DG units into the power network has many unknown variables, such as size and location of DG, number of customers who will utilize DG. The consumption profiles and output of DG are also intermittent in nature. These unknown variables have an impact on the HC. Considering the randomicity of some of these unknown variables, Probabilistic Power Flow (PPF) analysis is carried out. In PPF, several load flow calculations are performed for thousands of random cases of number, location and/or size of DG units. Networks' variables such as voltage, current, losses etc. are recorded against the performance limits for the determination of HC [73]. The stochastic HC estimation have been reported by many authors [74]-[80]. In [74], authors have proposed a stochastic multi-objective optimization model to maximize HC of the distribution network for wind power and minimize the energy procurement costs in a wind integrated power system, while considering technical and economic aspects. Cost of the purchased energy from upstream network, and operation and maintenance cost of wind farms are taken as objective functions. Analysis was carried out on a standard radial 69 bus distribution feeder and a practical 152 bus distribution system.

A detailed report on feeder modelling, analysis, and evaluation of issues to assess the impact of integration of distributed solar PV and to determine the hosting capacity prepared by EPRI is given in [75], [76]. An EPRI Technical Update report on developing a screening methodology, which effectively evaluates the new interconnection requests, while considering PV and feeder-specific factors is presented in [77]. This methodology has considered the peak load levels along with other critical factors, including PV location, aggregate PV effects, and most importantly specific feeder characteristics such as voltage class, voltage regulation schemes, and operating criteria. Authors in [78] have used a stochastic approach to simulate various DG deployment scenarios. Various limits such as over-voltages, voltage deviations and voltage unbalance are considered, while calculating HC. Le Baut et. al [ref] have presented a three-step probabilistic HC assessment technique in [79]. In [80], a high-resolution approach is presented to assess the DG and storage systems connection into the distribution network. Authors have used a stochastic demand model for this work.

Since HC is highly location based, integration of new DG can be accepted at some locations, but not at other places. The voltage profile of the feeders is one of the key criteria in defining the locational HC (LHC). Rylander et al [81] presented the concept of stream-lined methodology to calculate HC termed as stream-lined HC (SLHC). This method was developed by EPRI to calculate the HC of feeder considering size and location of DER, integration technology and physical characteristic of the feeders. Application of the results include improved screening tools, visualization of HC constrained feeders, and identification of problematic locations within a feeder. High penetration of DER/DG into the electrical networks adversely affects various performance indices. The primary issues used to identify hosting capacity include overvoltage, voltage deviation, element fault current, breaker reduction of reach, sympathetic breaker tripping, breaker/fuse coordination. Anti-islanding and large-scale PV protection issues with grounded wye-delta interconnect transformer have significant impact on HC, whereas voltage imbalance, overloads and harmonics have low impact on hosting capacity.

Determination of hosting capacity of distribution network can be broadly categorised as utility-based and customer-based. In utility-based, problem can be defined as an optimization problem with an objective to maximize the integration of DERs without technical violation in the distribution networks. In customer-based, stochastic methods are mostly used for assessment of HC, since in this case, utilities have no control over the number, locations, and size of DERs. There are four techniques used of the estimation of HC: (i) deterministic, (ii) stochastic, (iii) optimization-based, and (iv) stream-lined. Though, all these methods follow the same general procedure, but their implementation is quite different. In all the four methodologies, power flow analysis is carried out to find voltages and currents in the distribution networks. For power system simulation and analysis, a wide range of commercial and non-commercial software tools are also available; for examples, PSS Sincal Integrated Capacity Analysis Module [106], DigSILENT PowerFactory [107], NEPLAN [108], Synergi Electric [109], and CYME [110]. A list of these software programs is given in [111]. A brief and precise discussion on the methodology used for the assessment of HC and software tools is provided in [14] and references therein. A summary of the studies made by various researcher [82]-[105] is presented in Table 3.

Table 3

Studies that utilize various performance indices for assessment of hosting capacity

| Author/ Reference Year | DER Tech. | Performance Indices | Objective and Technique used | Remarks |
|---|---|---|---|---|
| Sakar et al, 2017 [82] | PV | Overvoltage, under voltage, current capacity of power lines, and harmonic distortion of the system | HC estimation of a distorted distribution system with Photovoltaic (PV)-based DG units, using optimization-based technique | Lower HC with higher distortion in distribution networks. |

| Sakar et al, 2018 [83] | PV | Bus voltage limits, line ampacities, and harmonic distortion limits | To find system's HC and parameters of the proposed filter, by an optimization algorithm. | HC decreases with the increase in utility side's background voltage distortion and load side's nonlinearity values. |
|---|---|---|---|---|
| Braga et al, 2018 [84] | PV | Harmonic distortion, and location | To find harmonic hosting capacity of the IEEE 13 bus network at four locations through sensitivity analysis on Matlab and OpenDSS. | Presence of capacitors lead to higher value of voltage total harmonic distortion THDv due to the parallel and series resonance. |
| Mirbagheri et al, 2018 [85] | DG | Steady-state voltage variations, rapid voltage changes, and thermal limits of lines and transformers | To determine HC of distribution grids even in case of uncertainties in grid parameters or lack of data, by using a novel method called 'Bricks approach. | The detailed topology of the grid and power profile for all the nodes are not required. Method is fast. |
| Chathurangi et al, 2018 [86] | PV | Voltage limits and network loss | To find technical impacts caused by higher penetration of roof top solar PVs on the operating performance of LV distribution | Further connections of solar will cause violation of acceptable voltage limits and increase network |

| | | | networks, by modelling the detailed network in DIgSILENT PowerFactory simulation platform. | losses due to the reverse power flow. Network already has 40% solar penetration level (based on transformer capacity) |
|---|---|---|---|---|
| Faishal Fuad et al, 2018 [87] | PV | Voltage violation, current violation and power flow direction | To find HC of the distribution network under different loading conditions, by an iterative method. | Study reveals that voltage violation occurs only when the load is less than 48% of the average value, while reverse power flow occurs in most of the cases. |
| Al-Saadi et al, 2018 [88] | PV | Overvoltage and over loading | Assessment of HC by conducting stochastic simulations, in which the uncertainties of the solar irradiance and load variation are considered. | Hourly assessment will help to serve the automation strategies, where corrective actions can be taken for HC maximization. Inclusion of more performance index, such as frequency variation index, may |

| | | | | further improve the method. |
|---|---|---|---|---|
| Peppanen et al, 2018 [89] | DER | Thermal limit, steady-state voltage and voltage deviation | To find HC of three real North American residential distribution systems. Load variability and unbalance are also analysed. | Results showed that the most significant limiting factor is voltage deviation. |
| Alturki et al, 2018 [90] | DG | Overvoltage and overloading | To estimate HC in distribution grids, by using an optimization-based method. | Low computational time as compared to traditional methods, while offering comparable results. |
| Abad et al, 2018 [91] | DG | Voltage deviation, load growth, network structure and type of DG | Probabilistic model based on a two-step linearization algorithm is presented to linearize the HC model. | HC can be approximated by a Gaussian-shape distribution. |
| Al-Saffar et al, 2019 [92] | PV | Overvoltage | HC of three real circuits in Alberta, Canada are evaluated using Monte Carlo simulation based probabilistic power flow method. | The use of stochastic approach can overcome the shortcomings of deterministic methods, and can, in principle, handle circuits of |

| | | | | any complexity. |
|---|---|---|---|---|
| Duwadi et al, 2019 [93] | PV | Overvoltage, undervoltage, and unbalance between the phases | To find HC of a feeder using Monte Carlo simulation. A parallel algorithm was developed using Message Passing Interface to decrease the simulation time. | PV penetration was found to be limited to 10%, on days when solar irradiance was high. PV penetration may go upto 50% on days with low energy demand. |
| Steyn et al, 2019 [94] | Roof-top PV | Voltage increase and equipment overload | To find impact on integration of distributed rooftop PV systems on three distribution networks (residential, commercial, and industrial) in Cape Town, South Africa. | Integration of PV systems decreases the line and transformer loading, but after a certain level, the effect of integration start to reverse due to increase in PV penetration level and reverse power flow. |
| Lillebo et al, 2019 [95] | EV | Overloading | To explore the impacts of increasing EV penetration level in a Norwegian distribution grid, by using real power measurements | Applying a fast charger in the grid with standard power factor of 0.98 lag caused significant voltage |

| | | | obtained from household smart meters in load flow analyses. | deviations at several locations, the worst of which reached close to 0.03 p.u |
|---|---|---|---|---|
| Soukaina et al, 2019 [96] | DG | Overvoltage, line overloading and transformer overloading | A new method to calculate HC of DG, considering the variety of distribution line configurations and cross admittance, is presented. The proposed method was also applied to an underground medium voltage grid. Simulation was carried out using Matlab and Etap. | Based on obtained results, it is shown that including line capacitances in calculation increases the maximum power of DG unit. |
| Ismael et al, 2019 [97] | PV | Bus voltage, line thermal capacity, power factor, and individual and total harmonic distortion | To find probabilistic hosting capacity (PHC) due to high penetration of PV units in a non-sinusoidal power distribution network, by using Monte Carlo simulation, considering various uncertain parameters, such as intermittent | Results shows that deterministic hosting capacity (DHC) studies, which ignore the uncertainty of electrical parameters, result in optimistic results that cause a noticeable underestimati |

| | | | output power of DGs, background voltage harmonics, load alteration, and filter parameters' variations. Hybrid PSO and gravitational search algorithm (PSOGSA) has been used for optimal design of the proposed filter. | on to the HC levels that are achieved from the probabilistic studies. |
|---|---|---|---|---|
| Abideen et al, 2019, [98] | PV | Bus overvoltage and distribution network losses | A new algorithm based on gradual increase in the PV penetration level is introduced to estimate the maximum PV penetration level in LV distribution networks using Matlab. | This technique can be applied to any distribution network, if the power demand and the solar irradiation data are available. |
| Li et al, 2019 [99] | DER | | A valuation scheme to quantify the value of DER in active distribution network (AND) is developed. A two-tier scheme is proposed to value and compensate DER portfolio | The proposed scheme will not only encourage the proactive investment of DERs in AND, but also help enhance the role of DERs in offering reliable, |

| | | | proposed by customers and independent third parties. | resilient, affordable and sustainable power supply to customers. |
|---|---|---|---|---|
| Essackjee et al, 2019 [100] | PV | System harmonics | A detailed study on maximum rooftop PV HC is presented. The level of harmonic pollution was observed with increasing penetration level of photovoltaic units and using different sizes of such units. | The findings can be used by an electrical utility to limit the number of rooftops photovoltaic being connected so as not to degrade the quality of supply. |
| Sadeghian et al, 2020 [101] | PV | Voltage violation and reverse power flow | An impact-assessment framework, based on Monte Carlo technique, to assess the impacts of two different types of distributed solar photovoltaic (DPV) installation (customer based and utility based) on a realistic distribution network. | To achieve higher penetration ratios without reverse power flow and other negative impacts, utility-aided installation is necessary. |
| Mulenga et al, 2021 [102] | PV | Overvoltage | A stochastic algorithm for estimation of HC of low-voltage network for solar | It is shown that performance indices used in |

| | | | PV is introduced. Two types of uncertainties aleatory and epistemic are distinctively considered. | deterministic and time-series HC studies need to be complemented with a planning risk percentile for stochastic study. |
|---|---|---|---|---|
| Du et al, 2021 [103] | PV-Wind-Load | Voltage violation | A stochastic framework for HC based on wind-PV-load temporal characteristic is presented from a probabilistic view. A discretization-aggregation method is introduced to generate filter extreme combinations. | Study helps in making better use of available renewable resources and promoting the application of distributed hybrid power generation in the power grid. |
| Paudyal et al, 2021 [104] | EV | Voltage limit, thermal limit | Study on EV HC by focusing on extreme fast charging (xFC) of EV charging loads is carried out to identify representative feeders of the utility distribution network in a certain region. | Study can help utilities plan for optimal system upgrades to facilitate future needs and greatly reduce the cost of EV integration. |
| Mulenga et al, 2021 [105] | EV | Undervoltage, transformer and feeder overload | A stochastic methodology to single-phase and | The method can be applied as an |

| | | | three-phase EV charge hosting capacity for distribution networks is developed. The method includes two types of uncertainties, aleatory and epistemic. | extension to find HC as a function of time of day, week, or year, and the best periods for charging can be identified and used in designing smart charging mechanisms. |
|---|---|---|---|---|

Enhancing hosting capacity of the network is considered as one of the important goals for distribution system operator (DSO). The technical solutions to enhance systems' HC have been grouped in three categories, such as DSO solutions, prosumer solution, and interactive solutions. The effectiveness of these solutions depends on various factors, such as technology readiness, investment cost, and impact on congestion and compliance with the applied grid codes. The various methods used are: reactive power control [112], [113], voltage control using on load tap changer (OLTC) [114], [115], [116], active power curtailment [117], [118], energy storage technologies [119], [120], network reconfiguration and reinforcement [121], and harmonic mitigation techniques [83], [122]. A brief study on the various techniques used for HC enhancement is presented in [11] and references therein. Authors in [112], [113] have reported that using reactive power control and energy storage system, DG penetration can be increased, and the system losses can be decreased. It is concluded that the model can be successfully utilized in smart grids, where integration of large-scale DG sources is preferred. Navarro et al [114] carried out a techno-economic study on the use of OLTC to mitigate overvoltage problem in UK caused by high PV penetration. Authors examined the local and remote voltage control methods and compared it with conventional reinforcement solutions. Rauma et al [115, 116] have studied the advanced control of OLTC transformers and its application in increasing HC. Authors developed an analytical model for a set of 631 real LV electrical systems in France, based on actual voltage measurements recorded by advanced metering tools, and concluded that the use of OLTC can enhance the HC. In [117], [118], authors presented a study on the role of active power curtailment and dynamic line rating in enhancing HC for LV and MV distribution networks. The authors have categorized the active power curtailment into soft and hard curtailments. Poulios in [119] studied the optimal size, location and economic aspects of battery energy storage systems (BESSs) for increasing the system HC and have stressed that the prices of BESS need to be greatly reduced in order to make it competitive to other available HC enhancement solutions. Authors in [120] proposed a cost-based multi-objective optimization tool built on Matlab platform to evaluate the BESS capacity. Authors examined the role of BESS for voltage regulation, network loss reduction and peak load reduction. Ismael et al [121] have examined the problem of selecting the optimal conductor for a real radial distribution network in Egypt and presented a novel feeder enforcement index to assist network planners and DSOs to recognize the feeders that need to be reinforced first. Bollen et al in [122] have discussed the HC assessment considering over and under voltage performance limits. Authors have also studied the impact of inter-harmonics and super-harmonics on HC estimation.

The key findings of the studies are summarized as follows [64]-[122]:

- Higher the percentage of time during which curtailment is acceptable, higher is the amount of production capacity that can be integrated to the network. However, even though the hosting capacity is increased, the proportion of additional energy that must be curtailed increases quickly making additional increase in hosting capacity less and less attractive.

- Hosting capacities for each feeder in the residential and commercial PV scenario and utility-scale PV scenario are different due to the possible PV locations. In the utility-scale analysis, the deployed PV could be located close to start-of-circuit at the substation or in feeder extremities. In the residential and commercial deployed PV scenario analysis, the PV location depends to a greater degree upon the customer location.

- The process of developing the baseline model and methodology for running the simulation, includes hundreds of different decisions, which can significantly shape the outcome, both in terms of final hosting capacity figure, and its accuracy, when compared to real life conditions.

- The streamlined methodology was significantly faster to run (from a computing standpoint), however it had accuracy issues (both over and underestimating hosting capacity) in a not-insignificant number of cases, particularly when it came to application on complex circuits and with respect to two of the four power system criteria that were evaluated: power quality/voltage and protection.

- The iterative methodology results were found to be sufficiently accurate, but running the model was computationally intense, and thus would require more resources to deploy and may not be able to be run as often as needed for the type of scenario analysis that may be used for planning.

- Based on the method for PV deployment, voltage imbalance typically decreases, and overloads seldom occur. This is strongly influenced by limiting the PV size to the customer peak load in the small-scale analysis. For balance large-scale PV, voltage imbalance seldom occurs, however the PV systems do have a slight potential to cause overloads.

- Deterministic method uses few input parameters that are readily available. It is fast and easily implementable and presents a quick overview of grid performance. It assumes fixed value of parameters and does not consider intermittent nature of RES. It does not consider uncertainties. HC obtained by deterministic method is an estimate of the worst-case scenario and not the true value. The impact, in this method, is overestimated and the hosting capacity underestimated.

- Stochastic technique considers uncertainties in the network and presents a realistic overview of the grid performance under renewable energy sources integration based on probability distribution functions or possibility theory. This method simulates realistic network scenarios and is less time consuming than time-series method. It accommodates all probabilistic distribution functions and is comparatively easy to execute.

- Stochastic method needs large computational time with the increase in uncertainties considered in large distribution networks. It does not assess the time-related operation of control elements and network performance, and requires use of probability distribution functions, which can affect the accuracy of the result. In this method, complexity increases with the increase in the number of uncertainty types, and the evaluation and interpretation of HC quantification becomes a difficult task.

- Time-Series method considers time correlation in the network, power consumption and production. Time varying impacts of renewable energy sources on the network and operation of control elements are considered in HC determination. The method presents realistic overview of the network performance, based on time varying nature of power consumption and production. This method can give information related to the 'when' and 'how' of the HC quantification.

- Time-Series method requires many measurements data (time series data), and a lot of simulation may be needed. Some performance indices may require very low resolution and pose a computational challenge. Also, method is very time consuming for high resolution simulations.

- Harmonic issues are strongly influenced by load and existing resonance rather than PV penetration.

- By injecting reactive power, larger loads like a fast charger or a large EV household charger might be installed in weaker parts of a power grid.

- Voltage deviation constraint does not affect the HC probability curve significantly, which implies that distributing DGs over the system decreases the importance of voltage deviation constraint.

- The voltage deviation constraint often limits the locational HC, but not the HC.
- Voltage unbalance factor (VUF), which is an important index for unbalanced systems does not constrain HC of the test system.
- DG technology has a great effect on probability curve of HC. This is probably, because the PV capacity factor is higher than wind capacity factor in the test system.
- Stochastic hosting capacity methods can also be applied for addressing the potential overload due to large amount of solar power.
- For single-phase units, installation of distributed energy storage can reduce the unacceptable voltage rise with the customer.
- Energy storage can provide reserves, change net load shape to minimize ramping requirements, and shift supply of variable generation to periods of increased net load.
- Voluntary load reduction or load shifting can provide multiple benefits to integrating solar and reducing curtailment, including reducing the dependence on partially loaded synchronous generators for providing frequency stability and operating reserves and changing the shape of net load, which can reduce ramp rates, better align solar supply with demand, and reduce peak capacity needs.
- Balancing supply and demand over larger areas reduces the net variability of load and renewable resources such as PV, owing to greater spatial diversity of variable generation resources.
- Changing the way the grid is scheduled and dispatched, including changes to market rules, does not require new technologies and often represents the 'least cost' way to aid vehicle to grid (VG) integration.
- Inverter-based solar and wind plants can provide the grid's frequency response needs as these plants become a larger proportion of the generation fleet and new mechanisms are developed.
- Generators can respond better to net load shape created by additional PV via increased ramp rates and ranges as well as the ability to start and stop more frequently.
- Due to the integration of distributed generation, the level of power quality disturbances may increase beyond what is acceptable for other customers.

### 6. Conclusion

The electricity technology sector has gone through a marked change from its traditional architecture of large-scale centralized electric power supply systems that take advantage of significant economies of scale. Renewable energy resources, particularly wind and solar energy in form of distributed generation certainly fit this trend. Thus, the traditional cost comparisons, based on large bulk energy market, may be very misleading. Distributed generation is likely to pioneer the development of a new power market, in which the new energy technology does not simply supply energy but must instead meet the demand of services like energy management, emergency or back-up power, environmental improvements and fuel diversity. Driven by the need for a more sustainable energy system, this new power production is often of the renewable type, connected through a power electronic converter interface. The excessive integration of distributed generation systems into the existing electrical networks may adversely affect the system's performance, and may lead to problems and operational limit violations, when the system exceeds its hosting capacity. Hence, it has now become essential to know the hosting capacity of the network to accept any further integration request. In this paper, a brief review on the concept of hosting capacity, techniques being used for its assessment and enhancement and operational limits is presented. Substantial progress has been made in HC assessment and enhancement covering analysis, simulation, and the hardware development and testing for efficiency maximization and cost minimization. However, many problems and issues, especially those related to the development of affordable, in-exhaustible and clean renewable energy technologies for huge longer-term benefits, and broad range of policies, market regulations, new grid codes and standards needed to be addressed for appropriate system planning and operation of the power system to supply a reliable and good quality electric power.

## 7. ACKNOWLEDGEMENTS

This work is a part of project ARMOUR supported by the European Union's Horizon 2020 program under the Marie Skłodowska-Curie grant agreement no. 890844.

## 8. REFERENCES

[1] M. Canale, I. Fagiano, M. KiteGene Milanese, 'A revolution in wind energy generation', Energy 2009; 34: 355-361.

[2] W. Tien, K-C. Kuo, 'An analysis of power generation from municipal solid waste (MSW) incineration plants in Taiwan', Energy 2010; 35: 4824-430.

[3] 'Solar energy perspectives: Executive summary', International Energy Agency; 2011.

[4] J-F. Mercure, P. Salas, 'An assessment of global energy resource economic potential', Energy 2012; 46: 322-336.

[5] R. W. Bacon, J. E. Besant-Jones, 'Global Electric Power Reform, Privatization and Liberalization of the Electric Power Industry in Developing Countries', Annual Reviews of Energy and the Environment 2002; 26: 331-359.

[6] H. Rudnick, J. Zolezzi, 'Electric sector deregulation and restructuring in Latin America: lessons to be learnt and possible ways forward', IEEE Proceedings Generation, Transmission and Distribution 2001; 148: 180-184.

[7] V. Foster, A. Rana, 'Rethinking power sector reform in the developing world', Sustainable Infrastructure 2020; Washington, DC: World Bank. © World Bank. License: CC BY 3.0 IGO

[8] P. N. Vovos, A. E. Kiprakis, A. R. Wallace, G. P. Harrison, 'Centralized and distributed voltage control: impact on distributed generation penetration', IEEE Trans. Power Systems 2007; 22: 476-483.

[9] M. Ebad, W. M. Grady, 'An approach for assessing high-penetration PV impact on distribution feeders', Electric Power System Research 2016; 133: 347-354.

[10] J. A. P. Lopes, N. Hatziargyriou, J. Mutale, P. Djapic, N. Jenkins, 'Integrating distributed generation into electric power systems: a review of drivers, challenges and opportunities', Electric Power System Research 2007; 77: 1189-1203.

[11] Sherif M. Smael, Shady H.E. Abdel Aleem, Almoataz Y. Abdelaziz, Ahmed F. Zobba, 'State-of-the-art of hosting capacity in modern power systems with distributed generation', Renewable Energy 2019; 130: 1002-1020.

[12] H. Bevrani, A. Ghosh, G. Ledwich, 'Renewable energy sources and frequency regulation: survey and new perspectives', IET Renewable Power Generation 2010; 4: 438-457.

[13] K. Dehghanpour, S. Afsharnia, 'Electrical demand side contribution to frequency control in power systems: a review on technical aspects', Renewable and Sustainable Energy Reviews 2015; 41: 1267-1276.

[14] Mohammad Zain ul Abideen, Omar Ellabban, Luluwah Al-Fagih, 'A review of the tools and methods for distribution networks' hosting capacity calculation', Energies 2020; 13: 2758.

36
[15] M.H.J Bollen, F. Hassan, 'Integration of distributed generation in the power system', Willey-IEEE Press: Hoboken, NJ, USA, 2011; pp 1-5.

[16] E. Mulenga, M.H.J Bollen, N. Etherden, 'A review of hosting capacity quantification methods for photovoltaics in low-voltage distribution grids', Int. J. Electric Power and Energy Systems 2020; 115: 105445.

[17] T. Stetz, W. Yan, M. Braun, 'Voltage control in distribution systems with high level penetration improving absorption capacity for PV system by reactive power supply', In Proceedings of the 25th Europian Photovoltaic Solar Energy Conference and Exhibition 2010; 49: pp. 1-7.

[18] N.E.M. Etherden, M.H.J. Bollen, 'Increasing the hosting capacity of distribution network by curtailment of renewable energy resources', In Proceedings of the 2011 IEEE Trondhei, PowerTech, Norway, 19-23 June 2011; pp. 1-7.

[19] John Kabouris, Fotis D. Kanellos, 'Impacts of Large Scale Wind Penetration on Energy Supply Industry', Energies 2009; 2: 1031-1041.

[20] J. P. Antoine, A. van Ranst, E. Stubbe, K. Derveaux, N. Janssens, H. Martinge, S. Vitet, N. M. Jensen, M. Durstewitz, J. Kabouris, D. V. Kanellopoulos, H. Bindner, 'IRENE 2010: Integration of the Renewable Energy in the Electrical Network', In Proceedings of the ALTENER 2000 Conference, Toulouse, France, 23–25 October, 2000.

[21] I. Erlich, W. Winter, A. Dittrich, 'Advanced Grid Requirements for the Integration of Wind Turbines into the German Transmission System', In Proceeedings of the IEEE Power Engineering Society General Meeting, Montreal, Canada, 18–22 June, 2006.

[22] F. Kanellos, N. Hatziargyriou, 'Control of variable speed wind turbines in islanded mode of operation', IEEE Transactions on Energy Conversion Journal 2008; 23: 535-543.

[23] M. Albadi, E. El-Saadany, 'Overview of wind power intermittency impacts on power system', Electric Power System Research 2010; 80 (6): 627-632.

[24] J. Smith, R. Thresher, R. Zavadil, E. DeMeo, R. Piwko, B. Ernst, T. Ackermann, 'A mighty wind', IEEE Power and Energy Magazine 2009; 7: 41-51.

[25]   J. Wang, A. Botterud, R. Bessa, H. KEKO, l. Carvalho, D. Issicaba, J. Sumaili, V. Miranda, 'Wind power forecasting uncertainty and unit commitment', Applied Energy 2011; 88(11): 4014-4023.

[26]   C. Lowery, M. O'Malley, 'Impact of wind forecast error statistics upon unit commitment', IEEE Transaction on Sustainable Energy 2012; 3 (4): 760-768.

[27]   Y. Zhang, J. Wang, X. Wang, 'Review of probabilistic forecasting of wind power generation', Renewable and sustainable energy reviews 2014; 32: 255-270.

[28]   C. Wan, Z. Xu, P. Pinson, Z. Y. Dong, K. P. Wong, 'Optimal prediction intervals of wind power generation', IEEE Transaction on power systems 2014; 29 (3): 1166-71174.

[29]   Y. Huang, Q. P. Zheng, J. Wang, 'Two-stage stochastic unit commitment model including non-generation resources with conditional value-at-risk constraints', Electric Power Systems Research 2014; 116: 427-438.

[30]   P. Li, X. Guan, J. Wu, 'Aggregated wind power generation probabilistic forecasting based on particle filter', Energy Conversion and Management 2015; 96: 579-587.

[31]   R. Chen et al., 'Reducing generation uncertainty by integrating CSP with wind power: An adaptive robust optimization-based analysis', IEEE Transactions on Sustainable Energy 2015; 6 (2): 583–594.

[32]   B. Hu, L. Wu, X. Guan, F. Gao, Q. Zhai, 'Comparison of variant robust SCUC models for operational security and economics of power systems under uncertainty', Electric Power Systems Research 2016; 133: 121–131.

[33]   M. Dreidy, H. Mokhalis, Saad Mekhilef, 'Inertia response and frequency control techniques for renewable energy sources: A review', Renewable and Sustainable Energy Reviews 2017; 69: 144-155.

[34]   X. Zheng, H. Chen, 'Data-driven distributionally robust unit commitment with Wasserstein metric: Tractable formulation and efficient solution method', IEEE Transactions on Power System 2020; 35 (6): 4940-4943.

[35] Y. Cho, T. Ishizaki, J-I Imura, 'Three-stage robust unit commitment considering decreasing uncertainty in wind power forecasting', IEEE Transactions on Industrial Informatics 2022; 18 (2): 796-806.

[36] G. K. Singh, 'Solar power generation by PV (photovoltaic) technology: a review', Energy 2013; 53: 1-13.

[37] Md. Shahariar Chowdhury, Kazi Sajedur Rahman, Tanzia Chowdhury et al., 'An overview of solar photovoltaic panels' end-of-life material recycling', Energy Strategy Reviews 2020; 27: 100431.

[38] 'Solar - Fuels & Technologies', International Energy Agency, Retrieved 18 June 2020.

[39] 'China: cumulative installed solar power capacity 2019', Statista, Retrieved 18 June 2020.

[40] 'Chinese Solar Perseveres During Pandemic', CleanTechnica, 21 May 2020. Retrieved 18 June 2020.

[41] 'IEA: Global Installed PV Capacity Leaps to 303 Gigawatts', greentechmedia, Eric Wesoff, 27 April 2017.

[42] 'Solar PV – Analysis', IEA, Retrieved 18 June 2020.

[43] 'IEA_PV_Snapshot_2020.pdf', International Energy Agency, Retrieved 2 May 2020.

[44] 'IEA_PV_Snapshot_2019.pdf', International Energy Agency, Retrieved 2 May 2020.

[45] 'Snapshot 2020 – IEA-PVPS', iea-pvps.org, Retrieved 10 May 2020.

[46] 'Renewable Capacity Statistics 2020', irena.org, Retrieved 23 May 2020.

[47] 'Snapshot 2021', IEA-PVPS, International Energy Agency.

[48] 'Renewable Capacity Statistics 2021' (PDF), irena.org, Retrieved 9 April 2021.

[49] J. Sterling, J. Mclaren, K. Cory, 'Treatment of solar generation in electric utility resource planning', Technical Report NREL/TP-6A20-60047, October 2013.

[50] K. N. Nwaigwe, P. Mutabilwa, E. Dinwa, 'An overview of solar power (PV systems) integration into electricity grids', Materials Science for Energy Technologies 2019; 2: 629-633.

[51] 'Investigating the impact of solar variability on grid stability', prepared by CAT Projects & ARENA (Australian Renewable Energy Agency) for public distribution, March 2015.

[52] P. P. Barker, R. W. de Mello, 'Determining the impact of distributed generation on power systems: Part I - radial distributed systems', Proceedings of 2000 IEEE PES Summer Meeting; 3: 1645-1656.

[53] K. L. Butler-Purry, M. Marotti, 'Impact of distributed generators on protective devices in radial distribution systems', 2005/2006 IEEE PES Transmission and Distribution Conference 2006; pp. 87-88.

[54] Y-K Wu, C-S Chen, Y-S Huang, C-Y lee, 'Advanced analysis of clustered photovoltaic system's performance based on the battery-integrated voltage control algorithm', International Journal of Emerging Electric Power Systems 2009; 10.

[55] M. Karimi, H. Mokhlis, K. Naidu, S. Uddin, A.H.A. Bakar, 'Photovoltaic penetration issues and impacts in distribution network – a review', Renewable and Sustainable Energy Reviews 2016; 53: 594-605.

[56] Chandra, A., Singh, G. K, and Pant, V., 'Protection of AC Microgrid Integrated with Renewable Energy Sources - A Research Review and Future Trends', Electric Power Systems Research, Vol. 179, April 2021, p. 107036.

[57] R. Albarracin, H. Amaris Duarte, 'Power quality in distribution power network with photovoltaic energy sources', 2009.

[58] J. Wang, Y. S. Lim J. H. Tong, E. Morris, 'Grid-connected photovoltaic system in Malaysia: a review on voltage issues', Renewable and Sustainable Energy Reviews 2014; 29: 535-545.

[59] F. Shania, A. Ghosh, G. Ledwich, F. Zare, 'Voltage unbalance improvement in low voltage resisdential feeders with rooftop PVs using custom power devices', International Journal of Electrical Power and Energy Systems 2014; 55: 362-377.

[60] A. Latif, D. Robinson, V. J. Gosbell, V. W. Smith, 'Harmonic impact of photovoltaic inverters on low voltage distribution system', 2006.

40
[61] 'Power quality and EMC issues with future electricity networks', Joint Working Group C4.24/CIRED, March 2018

[62] J. Leon, S. Kouro, L. Franquelo, J. Rodriguez, B. Wu, 'The Essential Role and the Continuous Evolution of Modulation Techniques for Voltage Source Inverters in Past, Present and Future Power Electronics', IEEE Transactions on Industrial Electronics 2016; 63 (5): 2688 – 2701.

[63] B. K. Bose, 'Multi-Level Converters', Electronics 2015; 4(3): 582-585.

[64] Nicholas Etherden, 'Increasing the Hosting Capacity of Distributed Energy Resources Using Storage and Communication', Doctoral Thesis, Lulea University of Technology, Sweden, 2014.

[65] M. H. J. Bollen and M. Häger, 'Power quality: interactions between distributed energy resources, the grid, and other customers', in First Int. Conf. on Renewable Energy Sources and Distributed Energy Resources, Brussels, 2004.

[66] G. P. Harrison, A. Piccolo, P. Siano and A. R. Wallace, 'Hybrid GA and OPF evaluation of network capacity for distributed generation connections', Electric Power Systems Research, vol. 78, no. 3, pp. 392-398, 2008.

[67] D. Menniti, M. Merlo, N. Scordino and F. Zanellini, 'Distribution network analysis: A comparison between hosting and loading capacities', in International Symposium on Power Electronics, Electrical Drives, Automation and Motion (SPEEDAM), Sorrento, 2012.

[68] M.H.J. Bollen and C. Coujard, 'Impact of small units for electricity generation', In 1st International Conference on Lifestyle, Health and Technology, Lulea, Sweden, June 2005.

[69] J. Deuse and G. Bourgain, editors, 'EU-DEEP Results: Integrating Distributed Energy Resources into Today's Electrical System', ExpandDER, www.expandDER.com, 2009.

[70] N. Etherden, M.H.J. Bollen, S. Ackeby, 'The transparent hosting-capacity approach – overview, applications and developments', in 23rd International Conference on Electricity Distribution, 15-18 June 2015.

[71]  B. Palmintier, R. Broderick, B. Mather, M. Coddington, K. Baker, F. Ding, M. Reno, M. Lave, A. Bharatkumar, 'On the path of SunShot: Emerging issues and challenges in integrating solar with the distribution system', 2016, NREL/TP-5D00-6533, SAND2016-2524R, NREL/TP-5D00-6531: SAND2016.

[72]  N. Etherden, M.H.J Bollen, 'Increasing the hosting capacity of distribution networks by curtailment of renewable energy resources', in 2011 IEEE Trondheim PowerTech, 19-23 June, 2011.

[73]  M. Rossi, G. Viganò, D. Moneta, D. Clerici, 'Stochastic evaluation of distribution network hosting capacity: Evaluation of the benefits introduced by smart grid technology', In Proceedings of the 2017 AEIT International Annual Conference, Cagliari, Italy, 20–22 September 2017; pp. 1–6.

[74]  A. Rabiee, S.M. Mohseni-Bonab, 'Maximizing hosting capacity of renewable energy sources in distribution networks: A multi-objective and scenario-based approach', Energy 2017; 120: 417-430.

[75]  M. Rylander, J. Smith, 'Stochastic analysis to determine feeder hosting capacity for distributed solar PV', EPRI Tech. Updat. 1026640 (2012), 1-50, 1026640.

[76]  'Alternative to the 15% rule: Modeling and hosting capacity analysis of 16 feeders', EPRI, Palo Alto, CA 2015, 3002005812.

[77]  S. Stanfield, 'IREC Series: Key lessons from California integrated capacity analysis', Interstate renewable energy council (IREC), 2017.

[78]  A. Dubey, S. Santoso, A. Maitra, 'Understanding photovoltaic hosting capacity of distribution circuits', in IEEE Power Energy Society General Meeting, 2015.

[79]  J. Le baut, P. Jehetbauer, S. Kadam, B. Bletterrie, N. Hatziargyriou, J. Smith, M. Rylander, 'Probabilistic evaluation of the hosting capacity in distribution networks', in IEEE PES Innov. Smart Grid Tech. Con. Eur, 2017.

[80]  E. J. Palacios-Garcia, A. Moreno-Munoz, I. Santiago, I.M. Moreo-Garcia, M.I. Milanes-Montero, 'PV hosting capacity analysis and enhancement using high resolution stochastic modeling, Energies 20017; 10:

[81]    M. Rylander, J. Smith, W. Sunderman, 'Streamline method for determining distribution system holding capacity', IEEE Trans. Industrial Applications 2016; 52: 105-111.

[82]    Sakar, S.; Balci, M.E.; Abdel Aleem, S.H.E.; Zobaa, A.F. Increasing PV hosting capacity in distorted distribution systems using passive harmonic filtering. Electr. Power Syst. Res. 2017, 148, 74–86.

[83]    Sakar, S.; Balci, M.E.; Abdel, S.H.E.; Zobaa, A.F. Integration of large- scale PV plants in non-sinusoidal environments: Considerations on hosting capacity and harmonic distortion limits. Renew. Sustain. Energy Rev. 2018, 82, 176–186.

[84]    Braga, M.D.; Machado, S.D.; Oliveira, I.C.; De Oliveira, T.E.C.; Ribeiro, P.F.; Lopes, B.I.L. Harmonic Hosting Capacity Approach in a Radial Distribution System due to PV Integration Using OpenDSS. In Proceedings of the 2018 13th IEEE International Conference on Industry Applications (INDUSCON), São Paulo, Brazil, 11–14 November 2018; pp. 222–228.

[85]    Mirbagheri, S.M.; Moncecchi, M.; Falabretti, D.; Merlo, M. Hosting Capacity Evaluation in Networks with Parameter Uncertainties. In Proceedings of the 2018 18th International Conference on Harmonics and Quality of Power (ICHQP), Ljubljana, Slovenia, 13–16 May 2018; pp. 1–6.

[86]    Chathurangi, D.; Jayatunga, U.; Rathnayake, M.; Wickramasinghe, A.; Agalgaonkar, A.; Perera, S. Potential Power Quality Impacts on LV Distribution Networks With High Penetration Levels of Solar PV. In Proceedings of the 2018 18th International Conference on Harmonics and Quality of Power (ICHQP), Ljubljana, Slovenia, 13–16 May 2018; pp. 1–6.

[87]    Faishal Fuad, R.S.; Adi, K.W.; Sarjiya; Putranto, L.M. Study on Photovoltaic Hosting in Yogyakarta Electric Distribution Network. In Proceedings of the 2018 5th International Conference on Information Technology, Computer, and Electrical Engineering (ICITACEE), Semarang, Indonesia, 27–28 September 2018; pp. 240–244.

[88]    Al-saadi, H.; Al-sarawi, S.; Zivanovic, R.; Abood, H.G. Hourly-Assessment of Grid Hosting Capacity for Active Distribution Network. In Proceedings of the

2018 IEEE International Conference on Probabilistic Methods Applied to Power Systems (PMAPS), Boise, ID, USA, 24–28 June 2018; pp. 1–7.

[89] Peppanen, J.; Bello, M.; Rylander, M. Service Entrance Hosting Capacity. In Proceedings of the 2018 IEEE 7th World Conference on Photovoltaic Energy Conversion (WCPEC) (A Joint Conference of 45th IEEE PVSC, 28th PVSEC & 34th EU PVSEC), Waikoloa Village, HI, USA, 10–15 June 2018; pp. 1451–1456.

[90] Alturki, M.; Khodaei, A.; Paaso, A.; Bahramirad, S. Optimization-based distribution grid hosting capacity calculations. Appl. Energy 2018, 219, 350–360.

[91] Abad, M.S.S.; Ma, J.; Zhang, D.; Ahmadyar, A.S.; Marzooghi, H. Probabilistic Assessment of Hosting Capacity in Radial Distribution Systems. IEEE Trans. Sustain. Energy 2018, 9, 1935–1947.

[92] Al-saffar, M.; Zhang, S.; Nassif, A.; Musilek, P. Assessment of Photovoltaic Hosting Capacity of Existing Distribution Circuits. In Proceedings of the 2019 IEEE Canadian Conference of Electrical and Computer Engineering (CCECE), Edmonton, AB, Canada, 5–8 May 2019; pp. 1–4.

[93] Duwadi, K.; Ingalalli, A.; Hansen, T.M. Monte Carlo Analysis of High Penetration Residential Solar Voltage Impacts using High Performance Computing. In Proceedings of the 2019 IEEE International Conference on Electro Information Technology (EIT), Brookings, SD, USA, 20–22 May 2019; pp. 1–6.

[94] Steyn, A.F.W.; Rix, A.J. Modelling the technical influence of randomly distributed solar PV uptake on electrical distribution networks. In Proceedings of the 2019 International Conference on Clean Electrical Power (ICCEP), Otranto, Italy, 2–4 July 2019; IEEE: Otranto, Italy, 2019; pp. 690–698.

[95] Lillebo, M.; Zaferanlouei, S.; Zecchino, A.; Farahmand, H. Impact of large-scale EV integration and fast chargers in a Norwegian LV grid. J. Eng. 2019, 2019, 5104–5108.

[96] Soukaina, N.; Hassane, E.; Hassan, E.M.; Tijani, L. Hosting capacity estimation of underground distribution feeder in Urbain Areas. In Proceedings of the 2019 International Conference on Wireless Technologies, Embedded and Intelligent Systems, WITS 2019, Fez, Morocco, 3–4 April 2019; pp. 1–5.

[97]    Ismael, S.M.; Aleem, S.H.E.A.; Abdelaziz, A.Y.; Zobaa, A.F. Probabilistic Hosting Capacity Enhancement in Non-Sinusoidal Power Distribution Systems Using a Hybrid PSOGSA Optimization Algorithm. Energies 2019, 12, 1018.

[98]    ul Abideen, M.Z.; Ellabban, O.; Refaat, S.S.; Abu-Rub, H.; Al-Fagih, L. A Novel Methodology to Determine the Maximum PV Penetration in Distribution Networks. In Proceedings of the 2019 2nd International Conference on Smart Grid and Renewable Energy (SGRE), Doha, Qatar, 19–21 Novermber 2019; pp. 1–6.

[99]    Li, Z.; Shahidehpour, M.; Alabdulwahab, A.; Al-Turki, Y. Valuation of distributed energy resources in active distribution networks. Electr. J. 2019, 32, 27–36.

[100]   Essackjee, I.A.; King, R.T.F.A. Maximum Rooftop Photovoltaic Hosting Capacity with Harmonics as Limiting Factor – Case Study for Mauritius. In Proceedings of the 2019 International Conference on Advances in Big Data, Computing and Data Communication Systems (icABCD), Winterton, South Africa, 5–6 August 2019; pp. 1–6.

[101]   Sadeghian, H.; Wang, Z. A novel impact-assessment framework for distributed PV installations in low-voltage secondary networks. Renew. Energy 2020, 147, 2179–2194.

[102]   E. Mulenga, M.H.J. Bollen, N. Etherden, 'Solar PV stochastic holding capacity in distribution networks considering aleatory and epistemic uncertainties', Electrical Power and Energy Systems 2021; 130: 106928.

[103]   N. Du, F. Tang, Q. Liao, C. Wang, X. Gao, J. Xie, J. Zhang, R. Lu, 'Hosting Capacity Assessment in Distribution Networks Considering Wind–Photovoltaic–Load Temporal Characteristics', Frontiers in Energy Research 2021, doi: 10.3389/fenrg.2021.767610.

[104]   P. Paudyal, S. Ghosh, S. Veda, D. Tiwari, J. Desai, 'EV Hosting Capacity Analysis on Distribution Grids', National Renewable Energy Laboratory (NREL) Report 2021.

45
[105] E. Mulenga, M.H.J. Bollen, N. Etherden, 'Adapted Stochastic PV Hosting Capacity Approach for Electric Vehicle Charging Considering Undervoltage', Electricity 2021; 2: 387-402.

[106] Siemens Maximal Hosting Capacity (ICA). Available online: https://assets.new.siemens.com/siemens/assets/public.1516636173.d30d49557176528d935ec035d8499ac26d083822.11-ica-module-datasheet-sincal-ag.pdf.

[107] DIgSILENT PowerFactory 2019 What's New. Available online: https://www.digsilent.de/en/downloads.html.

[108] NEPLAN Target Grid Planning. Available online: https://www.neplan.ch/description/target-grid-planning/

[109] Smarter Grid Solutions (SGS). Enhanced Hosting Capacity Analysis. 2018. Available online: http://mnsolarpathways.org/wp-content/uploads/2018/10/mn-solar-pathways_pv-hosting-capacity-report.pdf.

[110] CYME Integration Capacity Analysis. Available online: http://www.cyme.com/software/cymeica/

[111] Open Electrical Power Systems Analysis Software. https://wiki.openelectrical.org/index.php?title=Power_Systems_Analysis_Software.

[112] S. F. Santos, D.Z. Fitiwi, M. Shafie-Khah, A. W. Bizuayehu, C.M.P. Cabrita, J.P.S. Catalao', New multi-stage and stochastic mathematical model for maximizing RES hosting capacity- Part I:', IEEE Trans. Sustainable Energy 2017; 8: 304-319.

[113] S. F. Santos, D.Z. Fitiwi, M. Shafie-Khah, A. W. Bizuayehu, C.M.P. Cabrita, J.P.S. Catalao', New multi-stage and stochastic mathematical model for maximizing RES hosting capacity- Part II:', IEEE Trans. Sustainable Energy 2017; 8: 320-330.

[114] A. Navarro-Espinosa, LF. Ochoa, 'Increasing the PV hosting capacity of LV networks: OLTC-fitted transformers vs. reinforcements', in 2015 IEEE Power Energy Society Innov. Smart Grid Technol. Conf. ISGT, 2015.

[115] K. Rauma, F. Cadoux, N. Hadj-SaiD, A. Dufournet, C. Baudot, G. Roupioz, 'Assessment of the MV/LV on-load tap changer technology as a way to increase LV hosting capacity for photovoltaic power generators', in IET Conf. Proc. 2016.

[116] R. Kalle, 'Industrial aspects of voltage management and hosting capacity of photovoltaic power generation in low voltage network', Universite Grenoble Alpes, 2016, Ph.D. Thesis.

[117] N. Etherden, M.H.J. Bollen, 'Overload and overvoltage in low-voltage and medium-voltage networks due to renewable energy-some illustrative case studies', Electric Power Systems Research 2014; 114: 39-48.

[118] N. Etherden, M.H.J. Bollen, ' Increasing hosting capacity through dynamic line rating-risk aspects', in Cigre Int. Symposium –across Borders-HVDC Syst. Mark. Integre, 2015.

[119] V. Poulios, 'Optimal placement and sizing of battery storage to increase the PV hosting capacity of low voltage grids', ETH Zurich University, Zurich, Switzerland, 2014, M.Sc. Thesis.

[120] N. Jayasekara, M.A.S. Masoum, P.J. Wolf, 'Optimal operation of distributed energy storage systems to improve distribution network load and generation hosting capability', IEEE Trans. Sustainable Energy 2016; 7: 250-261.

[121] S.M. Ismael, S.H.E. Aleem, A.Y. Abdelaziz, A.F. Zobaa, 'Practical consideration for optimal conductor reinforcement and hosting capacity enhancement in radial distribution systems', IEEE Access 2018; 6: 27268-27277.

[122] M.H. j. Bollen, S.K. Ronnberg, 'Hosting capacity of the power grid for renewable electricity production and new consumption equipment', Energies 2017; 10: 1325.